\documentclass[letter]{aa} 
\usepackage{natbib}
\usepackage[varg]{txfonts}
\usepackage{graphicx}

\usepackage{lscape}
\usepackage{rotating}
\usepackage{multirow}
\usepackage{hhline}
\usepackage{tabularx}
\usepackage{booktabs}
\usepackage{caption}
\usepackage{pict2e,slashbox}

\usepackage[]{textpos}   
 \usepackage{color} 

\def\Msun{\ifmmode{\mathrm M_\odot}\else{M$_\odot$}\fi}

\usepackage{color}

\begin{document}


\title{Formation of S0 galaxies through mergers}
\subtitle{Explaining angular momentum and concentration change from spirals to S0s}

\author{M.~Querejeta\inst{1}
\and M.~C.~Eliche-Moral\inst{2}
\and T.~Tapia\inst{3}
\and A.~Borlaff\inst{4}
\and G.~van~de~Ven\inst{1}
\and M.~Lyubenova\inst{5}
\and M.~Martig\inst{1}
\and J.~Falc\'{o}n-Barroso\inst{4}
\and J.~M\'{e}ndez-Abreu\inst{6}
}

\institute{Max Planck Institute for Astronomy, K\"{o}nigstuhl 17, D-69117 Heidelberg, Germany,  \email{querejeta@mpia-hd.mpg.de}
  \and Departamento de Astrof\'{i}sica y Ciencias de la Atm\'{o}sfera, Universidad Complutense de Madrid, E-28040 Madrid, Spain
  \and Instituto de Astronom\'{i}a, Universidad Nacional Aut\'{o}noma de M\'{e}xico, BC-22800 Ensenada, Mexico
  \and Instituto de Astrof\'{i}sica de Canarias, C/ V\'{i}a L\'{a}ctea, E-38200 La Laguna, Tenerife, Spain
  \and Kapteyn Astronomical Institute, University of Groningen, Postbus 800, NL-9700 Groningen, The Netherlands
 \and School of Physics and Astronomy, University of St Andrews, North Haugh, St Andrews, KY16 9SS, UK}

\date{Received ..... / Accepted .....}

\abstract {The CALIFA team has recently found that the stellar angular momentum and concentration of late-type spiral galaxies are incompatible with those of lenticular galaxies (S0s), concluding that fading alone cannot satisfactorily explain the evolution from spirals into S0s. 
Here we explore whether major mergers can 
provide an alternative way to transform spirals into S0s by analysing
the spiral-spiral major mergers from the GalMer database that lead to realistic, relaxed S0-like galaxies.
We find that the change in stellar angular momentum and concentration can explain the differences in the $\lambda_\mathrm{Re}$--$R_{90}/R_{50}$ plane found by the CALIFA team.
Major mergers thus offer a feasible explanation for the transformation of spirals into S0s.
}


\titlerunning{Origin of S0s: Disentangling major mergers from fading.}
\authorrunning{Querejeta et al.}

\maketitle

\section{Introduction}
\label{Sec:introduction}

A kinematic classification of early-type galaxies (ETGs) might more closely reflect their true physical nature than photometry-based morphological classifications, which can be biased by inclination effects. In this sense, the ratio between ordered and random motion, $V/\sigma$, has long been used as a proxy for the rotational support of a galaxy: disc-like, rotation-dominated galaxies are intuitively associated with the highest $V/\sigma$, while the lowest values are expected in spheroidal-like, dispersion-dominated galaxies. With the advent of integral field spectroscopy, the SAURON and ATLAS$^\mathrm{3D}$ teams developed an improved parameter, $\lambda_\mathrm{Re}$, which divides galaxies into two groups: fast and slow rotators \citep[FRs and SRs, respectively;][]{2007MNRAS.379..401E,2011MNRAS.414..888E,2011MNRAS.413..813C}. Based on this kinematic definition, most spirals and lenticulars (S0s) are classified as FRs, while ellipticals are found in both groups. 

In the quest to understand the origin of FRs and SRs, a number of studies have focused on simulations.
\citet{2009MNRAS.397.1202J}
showed that $\lambda_\mathrm{Re}$ is a good indicator of the true stellar angular momentum in ETGs, and studied their possible merger origin, including the effect of gas. They already stated that the mass ratio of the encounter is crucial to determining the outcome as a SR or FR.
This is in agreement with studies pointing to equal-mass mergers as the most likely origin of elliptical galaxies 
\citep{2003ApJ...597..893N,2005ApJ...620L..79S,2010ApJ...723..818H,2011MNRAS.417..863D,2015ApJ...802L...3T}.
The results from \citet{2011MNRAS.416.1654B} also support this idea, and underline the role of spin (within 1:1 and 1:2 mergers, SRs are associated with retrograde and FRs with prograde encounters). These authors study the incidence of kinematic misalignments and kinematically decoupled cores, which are found primarily in SRs; however, they cannot explain the round SRs observed in ATLAS$^\mathrm{3D}$. A similar study with an exhaustive exploration of the parameter space was undertaken by \citet{2014MNRAS.444.1475M}, who conclude that round SRs can emerge from the accumulation of many minor mergers, while FRs may be formed through a variety of pathways. The idea that ETGs might have been sculpted through multiple mergers is not new \citep{1985MNRAS.215..517B, 1996ApJ...460..101W, 2001ApJ...546..189B, 2007A&A...476.1179B},
 and in fact different studies have analysed the effects of
minor and intermediate mergers onto early-type progenitors \citep[see e.g.][]{2004A&A...418L..27B,2009A&A...501L...9D,2011A&A...533A.104E,2013MNRAS.429.2924H,2015A&A...575A..16M,2015arXiv150407483Z}.
Using mock kinematic maps extracted from cosmological simulations, \citet{2014MNRAS.444.3357N} have recently confirmed that, even if the formation histories can be complex, the main results in a cosmological context are in full agreement with the conclusions drawn from idealised mergers.

As a consequence of the kinematic classification of ETGs,
it has become apparent that most S0s are FRs and have kinematic properties that make them
comparable to spirals \citep{2011MNRAS.416.1680C, 2011MNRAS.414.2923K}. Together with the fact that S0s span a whole range of bulge-to-disc ratios, this led \citet{2010MNRAS.405.1089L},
\citet{2011MNRAS.416.1680C}, and \citet{2012ApJS..198....2K} to propose a sequence of S0s parallel to that of spirals (i.e. S0a -- S0b -- S0c).
This recovers the original idea of \citet{1951ApJ...113..413S} and \citet{1976ApJ...206..883V}, bringing back the question of whether the classification parallelism reflects an underlying physical connection: \textit{are S0s faded spirals?}

Observations show that gas stripping, e.g. from ram pressure in clusters, can effectively transform spirals into S0s \citep{2005AJ....130...65C,2006A&A...458..101A,2008AJ....136.1623C, 
 2015MNRAS.447.1506M}. This constitutes an example of fading (any processes resulting in the suppression of star formation),
and it can contribute to explaining the observational morphology-density relation in high-density regimes \citep[][but see Cappellari et al. 2011]{1980ApJ...236..351D, 1997ApJ...490..577D}.

However, we know that S0s do not preferentially inhabit the densest cluster environments; they are equally common in groups \citep{2009ApJ...692..298W, 2011MNRAS.415.1783B}, where mergers and tidal effects dominate \citep{2014ApJ...782...53M,2014AdSpR..53..950M}. Additionally, traces of past mergers have been observed in some S0s \citep[e.g.][]{2004MNRAS.350...35F}. 
Therefore, it seems compelling to study under which conditions S0s could emerge out of mergers, and their impact on the observed kinematics.

The CALIFA team has recently raised further doubts about S0s as faded spirals through a new diagnostic diagram: the $\lambda_\mathrm{Re}$--concentration plane (van de Ven et al. in prep.). The population of late-type spirals (Sb, Sc, Sd) shows a clear incompatibility with S0s when both angular momentum ($\lambda_\mathrm{Re}$) and concentration ($R_{90}/R_{50}$) are simultaneously taken into account. 
 Provided that simple fading is not expected to significantly change the angular momentum of the galaxy, this contradicts the idea that \textit{most} S0s are faded spirals. Here, we study
whether major mergers can explain those differences using \textit{N}-body dissipative simulations from the GalMer database.
 We will analyse the encounters that end up in relaxed S0-like remnants, showing that the induced changes in ellipticity, stellar angular momentum, and concentration are in agreement with the CALIFA observations.

\section{Binary merger models} 
\label{Sec:models}

GalMer\footnote{GalMer project: http://galmer.obspm.fr} is a public database of binary $N$-body merger simulations that sample a wide range of mass ratios,
morphological types, 
and orbital characteristics. The progenitor galaxies are modelled using 
a spherical non-rotating dark-matter halo, an optional disc and bulge, with a spatial resolution of 0.28\,kpc and $\sim 10^5$ particles per galaxy.
The simulations use a TreeSPH code \citep{2002A&A...388..826S}, and take into account the effects of gas and star formation, with total simulation times in the range 3-4\,Gyr, which typically implies $\sim 1$\,Gyr of relaxation after full merger \citep[see][for more details]{2010A&A...518A..61C}.

\subsection{Lenticular remnants}
\label{Sec:lenticularem}

We consider the major mergers (mass ratios 1:1 to 1:3) involving all possible combinations of two spiral progenitors (Sa, Sb, or Sd) that give rise to realistic, dynamically-relaxed S0-like remnants, based on quantitative criteria that impose structural, kinematic, SFR, and gas-content parameters typical of 
lenticular galaxies (Eliche-Moral et al., in prep.).
In addition,
we have performed a visual morphological classification to identify which remnants would have been classified as S0-like by observers (i.e. as disc galaxies without noticeable spiral arms).
To this end, we have simulated photometric images of the resulting galaxies in several broad bands ($B$, $V$, $R$, $I$, and $K$), mimicking typical conditions of current observational surveys. 
We use a mass-to-light ratio which considers the stellar mass, age, and metallicity of each simulation particle according to \citet{2003MNRAS.344.1000B}, with a Chabrier IMF and the Padova evolutionary tracks; for details, see \citet{2014A&A...570A.103B} and \citet[][]{2015A&A...573A..78Q}.
 This provides us with a final sample of 67 S0-like remnants, which will be the focus of this letter.

\subsection{Stellar angular momentum, ellipticity, and concentration}
\label{Sec:lambda}

We assume that the simulated merger remnants are observed at the median distance of the 300 CALIFA galaxies in Falc{\'o}n-Barroso et al. (in prep.): $D_{\mathrm{median}} = 67$\,Mpc. We also consider the same spatial resolution and field of view as the survey, PSF$_\mathrm{FWHM}= 1.6"$, $R_{\mathrm{max}}=35"$ \citep{2012A&A...538A...8S}.

We calculate $\lambda_e$ according to \citet{2011MNRAS.414..888E},

\begin{equation}
\lambda_e = \frac{\sum F_i R_i |V_i|}{\sum F_i R_i \sqrt{V_i^2+\sigma_i^2}},
\end{equation}

\noindent
where $F_i$, $R_i$, $V_i$, and $\sigma_i$ are the flux (using an appropriate mass-to-light ratio), radius, velocity, and velocity dispersion measured within each spatial bin 
out to the effective radius of the galaxy ($R \leq R_\mathrm{e}$). The ellipticity at the effective radius $\varepsilon_\mathrm{e}$ is obtained from interpolation of the curve of growth of the luminosity-weighted ellipticities within increasingly larger isophotes.

The Petrosian concentration parameter $R_{90}/R_{50}$ 
It is defined as the ratio of the radii enclosing 90\% and 50\% of the Petrosian flux, measured on the 1D azimuthally averaged radial surface brightness profile in the simulated SDSS $r$ band. According to \citet{2009MNRAS.393.1531G}, there is an equivalence between $R_{90}/R_{50}$ and bulge-to-total ratio: $R_{90}/R_{50} = 1.93+2.02 \mathrm{(B+bar)/T}$, which we add on top of our plot for reference. This relation is
valid for $\mathrm{(B+bar)/T<0.6}$, but then flattens for higher values; thus, the upper axis in the right panel of Fig.\,1 must be interpreted with caution
for $\mathrm{(B+bar)/T>0.6}$.

\section{Results}
\label{Sec:results}

\begin{figure*}[t]
\begin{center}$
\begin{array}{ccc}
\includegraphics[trim=30 62 33 0,clip,height=0.45\textwidth]{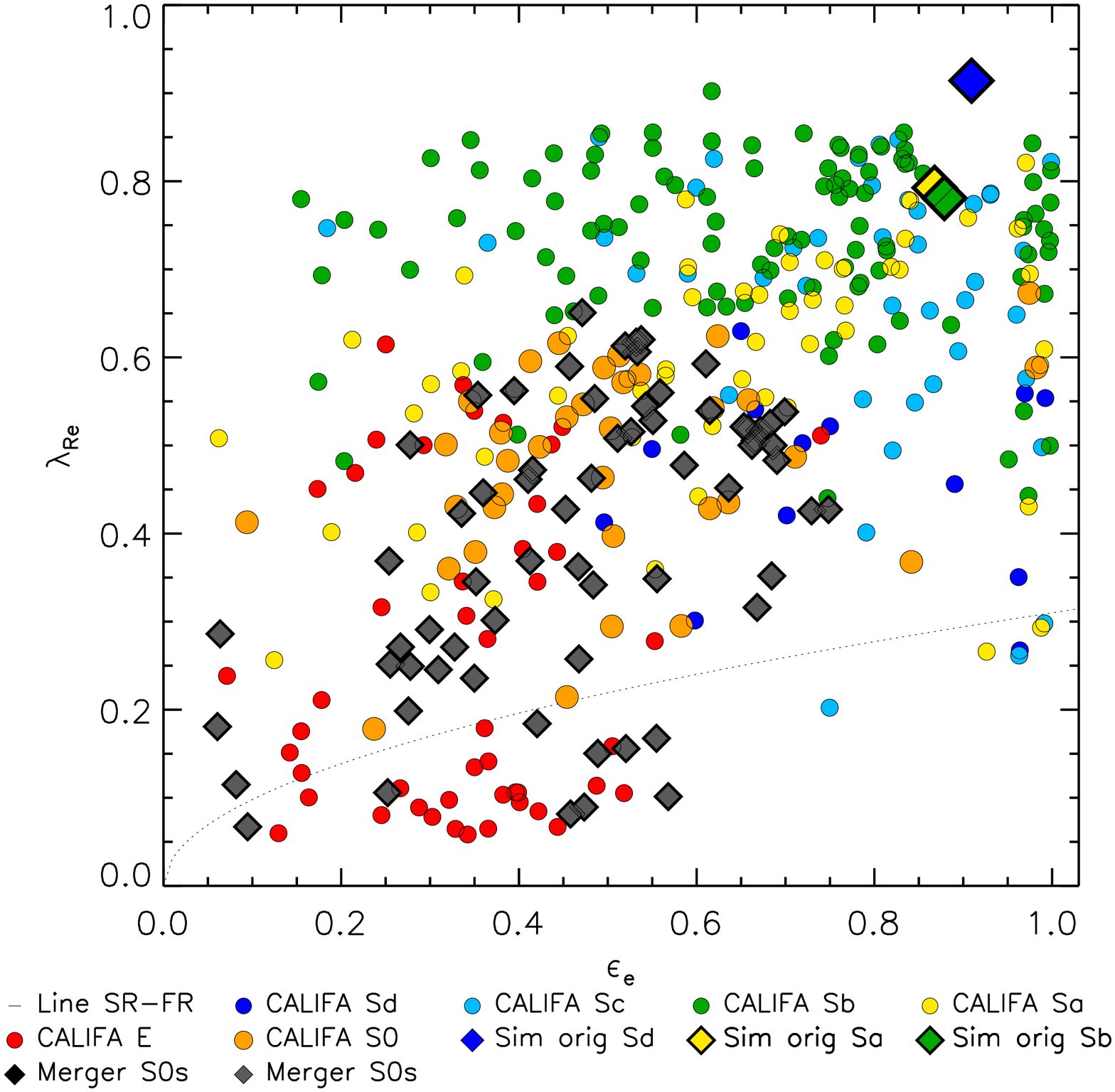} \hspace*{-0.9em} &
\includegraphics[trim=105 62 0 0,clip,height=0.45\textwidth]{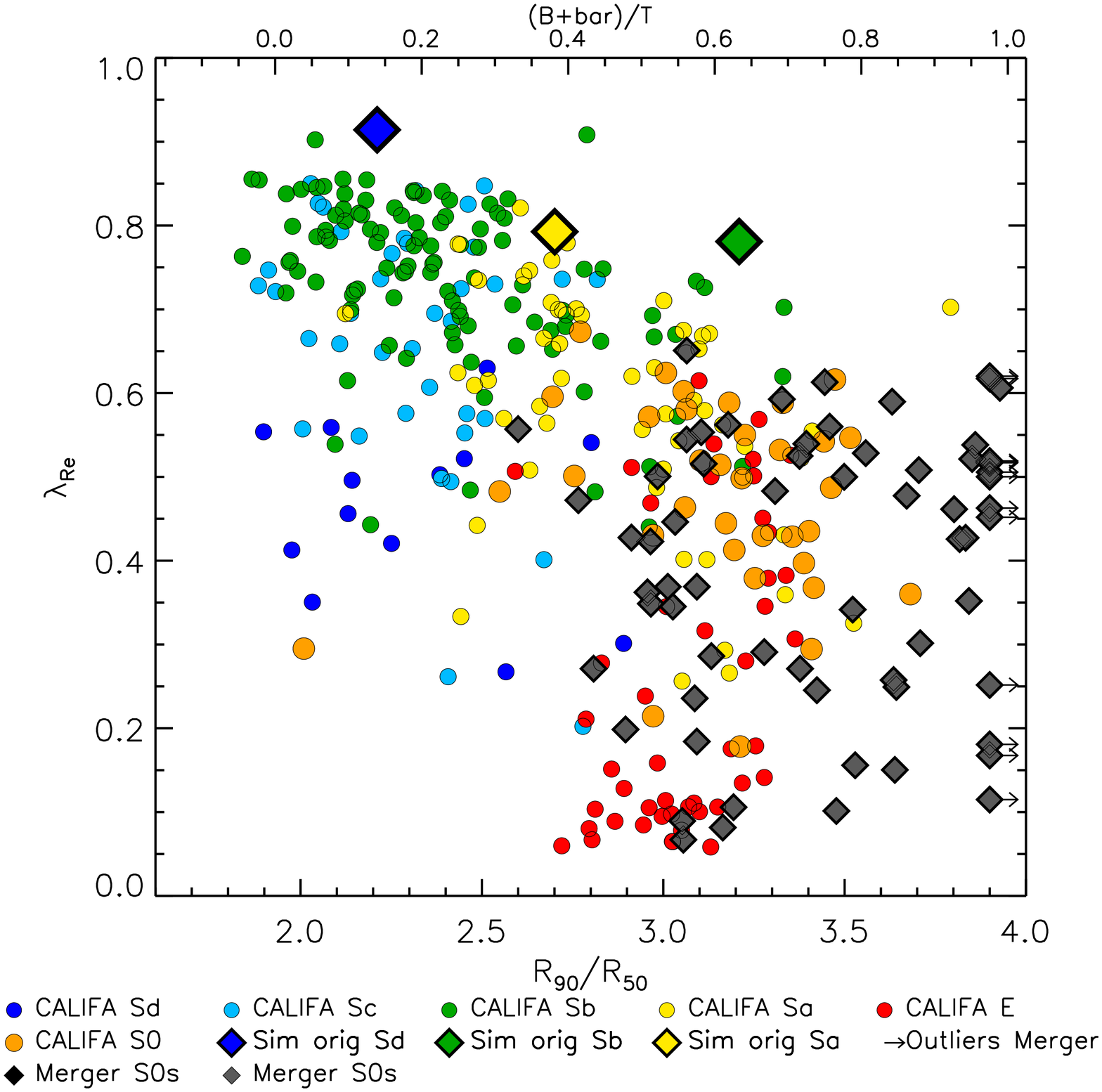} \hspace*{-1.0em}&
\includegraphics[trim=0 -50 0 -28,clip,height=0.45\textwidth]{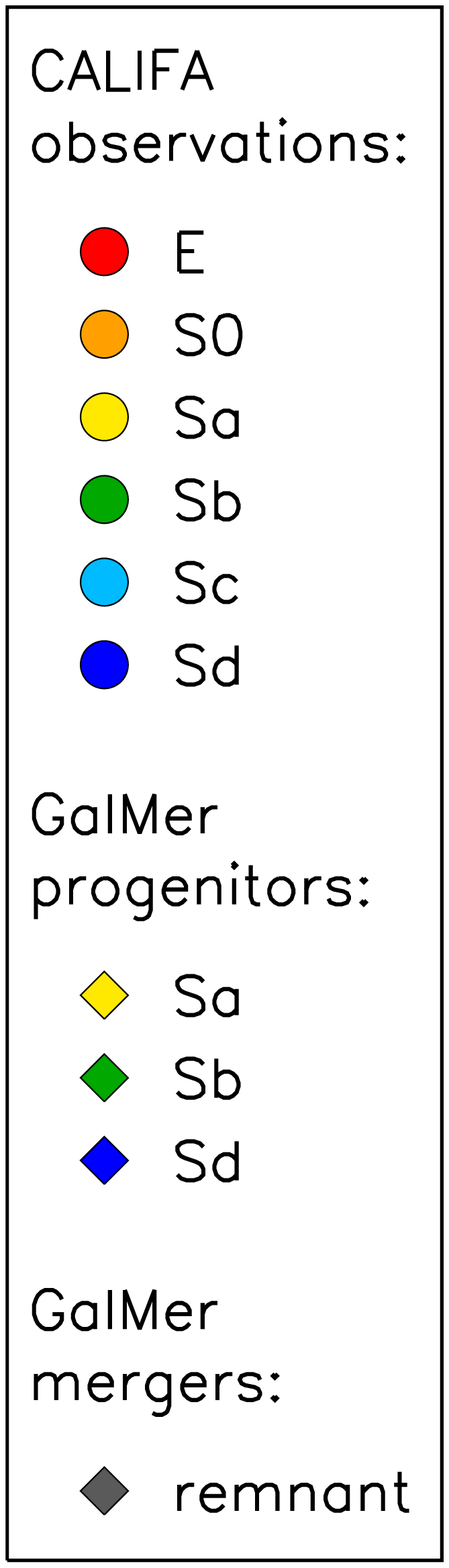}
\end{array}$
\caption{Stellar angular momentum ($\lambda_\mathrm{Re}$) plotted against the ellipticity ($\varepsilon_\mathrm{e}$, \textit{left}) and concentration ($R_{90}/R_{50}$, \textit{right}) 
for the GalMer simulations
 (original models and merger remnants), in comparison to CALIFA galaxies. 
All the parameters correspond to an edge-on view. 
The dotted line in the left plot represents the division between fast and slow rotators, and the top axis in the right plot is from \citet{2009MNRAS.393.1531G}, valid only up to $\mathrm{(B+bar)/T<0.6}$. See legend at the right for the symbols and colour-coding.}
\label{fig:lamepsconc}
\end{center}
\end{figure*}

\subsection{Angular momentum versus ellipticity}
\label{Sec:lambdaepsilon}

We compare the distribution of the remnants 
in the $\lambda_\mathrm{Re}$ -- $\varepsilon_\mathrm{e}$ plane with the real galaxies from the CALIFA survey. The spiral progenitor models lie in the region of FRs, as expected.

Naturally, for a meaningful comparison with observational data, one has to consider projection effects: the maximum $\lambda_\mathrm{Re}$ is attained for an edge-on view, and it is closer to the real angular momentum \citep{2009MNRAS.397.1202J}; for this reason, we compare our results with the CALIFA measurements \textit{corrected for inclination}. The deprojection of  $\lambda_\mathrm{Re}$ to edge-on follows van de Ven et al. (in prep.), assuming $\delta = 0.5$ for the global anisotropy (the mean value of the distribution proposed by Cappellari et al. 2007), but we check that varying it by $\pm 1 \sigma$ has a $< 3 \%$ effect on the resulting deprojected values. When deprojecting $\varepsilon_\mathrm{e}$,
due to observational uncertainties in the inclination, the result is unfeasible for a few cases, in which we assume the maximum $\varepsilon_\mathrm{intr} = 1$.

The left panel of Fig.~\ref{fig:lamepsconc} shows that major mergers can produce 
both fast- and slow-rotating S0s (9 out the 67 S0-like remnants are SRs). While only one SR S0s is present in the first CALIFA kinematic sample, in ATLAS$^\mathrm{3D}$, 13 out of the 36 SRs were S0-like \citep{2011MNRAS.413..813C}. We also find that, in agreement with previous studies, mergers tend to reduce both the stellar angular momentum $\lambda_\mathrm{Re}$ and ellipticity $\varepsilon_\mathrm{e}$ by an amount that varies largely depending on the mass ratio, gas fractions, and orbital characteristics  \citep{2009MNRAS.397.1202J,2011MNRAS.416.1654B,2014A&A...565A..31T,2014MNRAS.444.3357N}.
In this diagram, real spirals (especially Sb and Sc) have a very limited overlap with S0s. Remarkably, the S0-like remnants that we find in these simulations are in excellent agreement with the real S0s, and far from the progenitor spirals that they originate from. This is a first hint that a good fraction of S0s could have originated through a mechanism involving mergers.

\subsection{Angular momentum versus concentration}
\label{Sec:lambdaconc}

As mentioned in Sect.\,1, the differences between spirals and S0s become more striking in the $\lambda_\mathrm{Re}$--concentration plane introduced by van de Ven et al. (in prep.).
The right panel of Fig.\,\ref{fig:lamepsconc} compares the distribution of our remnant S0s with that of real galaxies from CALIFA in 
$\lambda_\mathrm{Re}$ versus $R_{90}/R_{50}$,
showing that S0s resulting from major mergers are consistent with 
real lenticulars.
In this plane, Sb and Sc galaxies clearly cluster towards the top-left corner; Sa galaxies show a larger scatter, extending down diagonally towards the bottom-right, where S0 and E galaxies are located. The number of Sd galaxies in this first CALIFA kinematic sample is small,
but they also show little overlap with S0s.

The novel and main result presented in this letter is that major mergers can transform spiral progenitors 
on the region of high $\lambda_\mathrm{Re}$ and low concentration into realistic S0 systems of lower $\lambda_\mathrm{Re}$ and higher concentration, in agreement with the CALIFA observations.
 The bulk shift between spirals and lenticulars is significant both observationally and according to these simulations. This confirms that \textit{major mergers are a plausible mechanism to explain this observed change, which is incompatible with simple fading}.

It might seem paradoxical that the progenitor Sa galaxy has a lower $R_{90}/R_{50}$ concentration than the Sb galaxy, but this should not be seen as unrealistic: in the CALIFA sample, many Sa indeed show  lower $R_{90}/R_{50}$ than a great deal of Sb galaxies. Even if B/T and concentration are often invoked as a proxy for the morphological type, there is not a one-to-one relation between them \citep[see e.g.][]{2008MNRAS.388.1708G}.

Within the S0 merger remnants there are some outliers, especially in terms of concentration, $R_{90}/R_{50}$, because for these remnants the total simulated time after coalescence is short (typically $\lesssim 0.5$\,Gyr after full merger). Since we consider a mass-to-light ratio (M/L) that varies according to the age and metallicity of the stellar particles in the simulation, when star formation takes place in the centre \citep[which is very frequent in mergers, see][]{1991ApJ...370L..65B}, we end up having an unusually high light concentration due to young stars, which reflects on the $R_{90}/R_{50}$ parameter \citep[see also][on the evolution of surface density profiles for elliptical galaxies resulting from simulated mergers]{2006MNRAS.369..625N,2009ApJS..181..135H}. In fact, the most extreme concentration outliers ($R_{90}/R_{50}$\,$> 4.5$) correspond to the encounters involving the highest initial gas fractions ($\gtrsim 20\%$), and the highest fractions of merger-triggered star formation (new stars $\gtrsim 15\%$ of total stellar mass).
 The variation of M/L in simulations is often overlooked (assuming a constant value), and this is another important result: if we want to measure concentrations for wet merger remnants, the effects of young stellar populations on the light distribution cannot be ignored, as already pointed out by \citet{2009MNRAS.397.1202J}. After an extra $\sim 1$\,Gyr of relaxation, the central young populations will decrease their $r$-band luminosity by at least a factor of 2 \citep[see][]{2015A&A...573A..78Q}, and this would probably shift the points to the left and fully overlap the observational values.
Additionally, it must be emphasised that most remnants have reasonable concentrations even after the short relaxation periods considered ($\sim 90$\% with $R_{90}/R_{50}<4$), fully compatible with the observed values for S0s.

The dependence on initial conditions is complex, but we have found a clear trend for the highest reduction in angular momentum to be associated with retrograde encounters, in agreement with \citet{2011MNRAS.416.1654B}. In fact, the emergence of SRs is almost only associated with retrograde mergers (8 out of 9). We have observed some degree of dependence with the pericentre distance, with a preference for short pericentres in the formation of SRs, but the dependence is milder than the correlation with the spin-orbit coupling. Within the range of major mergers considered, the mass ratio 1:1 to 1:3 does not seem to affect the outcome significantly. Interestingly, the change in $\lambda_\mathrm{Re}$ does not show any systematic dependence on the gas content or measured SFRs. In terms of concentration, however, there is some relation with the gas content and newborn stars, as commented above:
 the highest concentrations are preferentially associated with the most gas-rich encounters.

\section{Discussion}
\label{Sec:discussion}

As we have seen, the systematic offset between spirals and S0s in the $\lambda_\mathrm{Re}$--concentration plane reported by the CALIFA team can be attributed to the effect of major mergers. 
A number of studies have already made it clear that disc-like remnants can emerge from mergers, even those with  equal mass-ratios \citep[][but see the doubts raised by \citet{2011ApJ...730....4B} about the need to include supersonic turbulence in such simulations of gas-rich mergers]{2002MNRAS.333..481B, 2005ApJ...622L...9S, 2004ApJ...606...32R, 2006ApJ...641...90R, 2009ApJ...691.1168H, 2011MNRAS.415.3750M}. 
In this sense, in \citet[][]{2015A&A...573A..78Q} we proved that not only can discs survive major mergers, but also that they can  produce remnants with disc-bulge coupling in perfect agreement with observations of S0s \citep[e.g.][]{2010MNRAS.405.1089L}. Moreover, we have  found that the discs of our S0-like remnants tend to exhibit antitruncations in agreement with those in real S0s \citep[even reproducing their tight scaling relations, see][]{2014A&A...570A.103B}.
 If we combine these results with the angular momentum and concentration changes presented in this letter, 
 major mergers stand out as a plausible transformation mechanism to explain the origin of many S0s.
 
The role of mergers with other mass ratios (intermediate and minor mergers) and the combined effect of multiple mergers remains to be quantified. Based on the dry minor mergers considered by \citet{2014A&A...565A..31T}, and the studies by \citet{2011MNRAS.416.1654B}, \citet{2014MNRAS.444.1475M} and \citet{2014MNRAS.444.3357N}, we would expect intermediate and minor mergers to induce more modest changes on $\lambda_\mathrm{Re}$ and concentrations, but a series of mergers (which is very probable in a cosmological context) could add up and produce a net effect similar to the one that we have presented here. A major merger origin of a significant fraction of S0s would also be in agreement with the fact that higher gas fractions are expected in the early Universe, making it more likely to obtain disc-like remnants out of major mergers (keeping in mind  that the gas fractions of our progenitors are representative of present-day spirals).

The fact that our results are compatible with a merger origin of S0s should not, of course, be overinterpreted. We are claiming it as a \textit{plausible} mechanism, but not necessarily the only one, to explain the observed offset between most spirals and S0s in the $\lambda_\mathrm{Re}$--concentration plane. In fact, some spirals and S0s clearly overlap (especially Sa galaxies); in these cases, simple fading cannot be ruled out. Remarkably, from their analysis of pseudobulges, \citet{2015arXiv150307635V} have recently pointed out that gas stripping alone is a viable process to transform early-type spirals into S0s, but not for late-type spirals, which agrees with the overlap found by CALIFA.
 Thus, it is not reasonable to claim that all lenticular galaxies are the by-products of mergers, but their relative relevance is probably higher in the transformation starting from late-type spirals.
 Additionally, the role of merging in ``pre-processing'' galaxies in filaments before falling into a cluster and in ``post-processing'' them during their infall should also be considered \citep[see][]{2013MNRAS.435.2713V, 2014MNRAS.440.1690H}.

\small  
%
\begin{acknowledgements}   

The authors would like to acknowledge the referee for a helpful report, and the GalMer team for creating such a powerful tool, with especial thanks to Paola Di Matteo for her kind support with the database.
MQ, GvdV and JFB acknowledge financial support to the DAGAL network from the People Programme (Marie Curie Actions) of the European Union's Seventh Framework Programme FP7/2007- 2013/ under REA grant agreement number PITN-GA-2011-289313.
Supported by the Spanish Ministry of Economy and Competitiveness (MINECO) under projects AYA2012-31277 and AYA2013-48226-C3-1-P. JMA acknowledges support from the European Research Council Starting Grant (SEDmorph; P.I. V. Wild).
\end{acknowledgements}

\bibliographystyle{aa}
\bibliography{/Users/querejeta/Documents/E1_SCIENCE_UTIL/mq.bib}{}

\begin{thebibliography}{61}
\expandafter\ifx\csname natexlab\endcsname\relax\def\natexlab#1{#1}\fi

\bibitem[{{Arag{\'o}n-Salamanca} {et~al.}(2006){Arag{\'o}n-Salamanca},
  {Bedregal}, \& {Merrifield}}]{2006A&A...458..101A}
{Arag{\'o}n-Salamanca}, A., {Bedregal}, A.~G., \& {Merrifield}, M.~R. 2006,
  \aap, 458, 101

\bibitem[{{Barnes}(1985)}]{1985MNRAS.215..517B}
{Barnes}, J. 1985, \mnras, 215, 517

\bibitem[{{Barnes}(2002)}]{2002MNRAS.333..481B}
{Barnes}, J.~E. 2002, \mnras, 333, 481

\bibitem[{{Barnes} \& {Hernquist}(1991)}]{1991ApJ...370L..65B}
{Barnes}, J.~E. \& {Hernquist}, L.~E. 1991, \apjl, 370, L65

\bibitem[{{Bekki}(2001)}]{2001ApJ...546..189B}
{Bekki}, K. 2001, \apj, 546, 189

\bibitem[{{Bekki} \& {Couch}(2011)}]{2011MNRAS.415.1783B}
{Bekki}, K. \& {Couch}, W.~J. 2011, \mnras, 415, 1783

\bibitem[{{Bois} {et~al.}(2011){Bois}, {Emsellem}, {Bournaud}, {Alatalo},
  {Blitz}, {Bureau}, {Cappellari}, {Davies}, {Davis}, {de Zeeuw}, {Duc},
  {Khochfar}, {Krajnovi{\'c}}, {Kuntschner}, {Lablanche}, {McDermid},
  {Morganti}, {Naab}, {Oosterloo}, {Sarzi}, {Scott}, {Serra}, {Weijmans}, \&
  {Young}}]{2011MNRAS.416.1654B}
{Bois}, M., {Emsellem}, E., {Bournaud}, F., {et~al.} 2011, \mnras, 416, 1654

\bibitem[{{Borlaff} {et~al.}(2014){Borlaff}, {Eliche-Moral},
  {Rodr{\'{\i}}guez-P{\'e}rez}, {Querejeta}, {Tapia}, {P{\'e}rez-Gonz{\'a}lez},
  {Zamorano}, {Gallego}, \& {Beckman}}]{2014A&A...570A.103B}
{Borlaff}, A., {Eliche-Moral}, M.~C., {Rodr{\'{\i}}guez-P{\'e}rez}, C.,
  {et~al.} 2014, \aap, 570, A103

\bibitem[{{Bournaud} {et~al.}(2011){Bournaud}, {Chapon}, {Teyssier}, {Powell},
  {Elmegreen}, {Elmegreen}, {Duc}, {Contini}, {Epinat}, \&
  {Shapiro}}]{2011ApJ...730....4B}
{Bournaud}, F., {Chapon}, D., {Teyssier}, R., {et~al.} 2011, \apj, 730, 4

\bibitem[{{Bournaud} {et~al.}(2004){Bournaud}, {Combes}, \&
  {Jog}}]{2004A&A...418L..27B}
{Bournaud}, F., {Combes}, F., \& {Jog}, C.~J. 2004, \aap, 418, L27

\bibitem[{{Bournaud} {et~al.}(2007){Bournaud}, {Jog}, \&
  {Combes}}]{2007A&A...476.1179B}
{Bournaud}, F., {Jog}, C.~J., \& {Combes}, F. 2007, \aap, 476, 1179

\bibitem[{{Bruzual} \& {Charlot}(2003)}]{2003MNRAS.344.1000B}
{Bruzual}, G. \& {Charlot}, S. 2003, \mnras, 344, 1000

\bibitem[{{Cappellari} {et~al.}(2011{\natexlab{a}}){Cappellari}, {Emsellem},
  {Krajnovi{\'c}}, {McDermid}, {Scott}, {Verdoes Kleijn}, {Young}, {Alatalo},
  {Bacon}, {Blitz}, {Bois}, {Bournaud}, {Bureau}, {Davies}, {Davis}, {de
  Zeeuw}, {Duc}, {Khochfar}, {Kuntschner}, {Lablanche}, {Morganti}, {Naab},
  {Oosterloo}, {Sarzi}, {Serra}, \& {Weijmans}}]{2011MNRAS.413..813C}
{Cappellari}, M., {Emsellem}, E., {Krajnovi{\'c}}, D., {et~al.}
  2011{\natexlab{a}}, \mnras, 413, 813

\bibitem[{{Cappellari} {et~al.}(2011{\natexlab{b}}){Cappellari}, {Emsellem},
  {Krajnovi{\'c}}, {McDermid}, {Serra}, {Alatalo}, {Blitz}, {Bois}, {Bournaud},
  {Bureau}, {Davies}, {Davis}, {de Zeeuw}, {Khochfar}, {Kuntschner},
  {Lablanche}, {Morganti}, {Naab}, {Oosterloo}, {Sarzi}, {Scott}, {Weijmans},
  \& {Young}}]{2011MNRAS.416.1680C}
{Cappellari}, M., {Emsellem}, E., {Krajnovi{\'c}}, D., {et~al.}
  2011{\natexlab{b}}, \mnras, 416, 1680

\bibitem[{{Chilingarian} {et~al.}(2010){Chilingarian}, {Di Matteo}, {Combes},
  {Melchior}, \& {Semelin}}]{2010A&A...518A..61C}
{Chilingarian}, I.~V., {Di Matteo}, P., {Combes}, F., {Melchior}, A.-L., \&
  {Semelin}, B. 2010, \aap, 518, A61

\bibitem[{{Crowl} \& {Kenney}(2008)}]{2008AJ....136.1623C}
{Crowl}, H.~H. \& {Kenney}, J.~D.~P. 2008, \aj, 136, 1623

\bibitem[{{Crowl} {et~al.}(2005){Crowl}, {Kenney}, {van Gorkom}, \&
  {Vollmer}}]{2005AJ....130...65C}
{Crowl}, H.~H., {Kenney}, J.~D.~P., {van Gorkom}, J.~H., \& {Vollmer}, B. 2005,
  \aj, 130, 65

\bibitem[{{Di Matteo} {et~al.}(2009){Di Matteo}, {Jog}, {Lehnert}, {Combes}, \&
  {Semelin}}]{2009A&A...501L...9D}
{Di Matteo}, P., {Jog}, C.~J., {Lehnert}, M.~D., {Combes}, F., \& {Semelin}, B.
  2009, \aap, 501, L9

\bibitem[{{Dressler}(1980)}]{1980ApJ...236..351D}
{Dressler}, A. 1980, \apj, 236, 351

\bibitem[{{Dressler} {et~al.}(1997){Dressler}, {Oemler}, {Couch}, {Smail},
  {Ellis}, {Barger}, {Butcher}, {Poggianti}, \&
  {Sharples}}]{1997ApJ...490..577D}
{Dressler}, A., {Oemler}, Jr., A., {Couch}, W.~J., {et~al.} 1997, \apj, 490,
  577

\bibitem[{{Duc} {et~al.}(2011){Duc}, {Cuillandre}, {Serra}, {Michel-Dansac},
  {Ferriere}, {Alatalo}, {Blitz}, {Bois}, {Bournaud}, {Bureau}, {Cappellari},
  {Davies}, {Davis}, {de Zeeuw}, {Emsellem}, {Khochfar}, {Krajnovi{\'c}},
  {Kuntschner}, {Lablanche}, {McDermid}, {Morganti}, {Naab}, {Oosterloo},
  {Sarzi}, {Scott}, {Weijmans}, \& {Young}}]{2011MNRAS.417..863D}
{Duc}, P.-A., {Cuillandre}, J.-C., {Serra}, P., {et~al.} 2011, \mnras, 417, 863

\bibitem[{{Eliche-Moral} {et~al.}(2011){Eliche-Moral},
  {Gonz{\'a}lez-Garc{\'{\i}}a}, {Balcells}, {Aguerri}, {Gallego}, {Zamorano},
  \& {Prieto}}]{2011A&A...533A.104E}
{Eliche-Moral}, M.~C., {Gonz{\'a}lez-Garc{\'{\i}}a}, A.~C., {Balcells}, M.,
  {et~al.} 2011, \aap, 533, A104+

\bibitem[{{Emsellem} {et~al.}(2011){Emsellem}, {Cappellari}, {Krajnovi{\'c}},
  {Alatalo}, {Blitz}, {Bois}, {Bournaud}, {Bureau}, {Davies}, {Davis}, {de
  Zeeuw}, {Khochfar}, {Kuntschner}, {Lablanche}, {McDermid}, {Morganti},
  {Naab}, {Oosterloo}, {Sarzi}, {Scott}, {Serra}, {van de Ven}, {Weijmans}, \&
  {Young}}]{2011MNRAS.414..888E}
{Emsellem}, E., {Cappellari}, M., {Krajnovi{\'c}}, D., {et~al.} 2011, \mnras,
  414, 888

\bibitem[{{Emsellem} {et~al.}(2007){Emsellem}, {Cappellari}, {Krajnovi{\'c}},
  {van de Ven}, {Bacon}, {Bureau}, {Davies}, {de Zeeuw}, {Falc{\'o}n-Barroso},
  {Kuntschner}, {McDermid}, {Peletier}, \& {Sarzi}}]{2007MNRAS.379..401E}
{Emsellem}, E., {Cappellari}, M., {Krajnovi{\'c}}, D., {et~al.} 2007, \mnras,
  379, 401

\bibitem[{{Falc{\'o}n-Barroso} {et~al.}(2004){Falc{\'o}n-Barroso}, {Peletier},
  {Emsellem}, {Kuntschner}, {Fathi}, {Bureau}, {Bacon}, {Cappellari}, {Copin},
  {Davies}, \& {de Zeeuw}}]{2004MNRAS.350...35F}
{Falc{\'o}n-Barroso}, J., {Peletier}, R.~F., {Emsellem}, E., {et~al.} 2004,
  \mnras, 350, 35

\bibitem[{{Gadotti}(2009)}]{2009MNRAS.393.1531G}
{Gadotti}, D.~A. 2009, \mnras, 393, 1531

\bibitem[{{Graham} \& {Worley}(2008)}]{2008MNRAS.388.1708G}
{Graham}, A.~W. \& {Worley}, C.~C. 2008, \mnras, 388, 1708

\bibitem[{{Head} {et~al.}(2014){Head}, {Lucey}, {Hudson}, \&
  {Smith}}]{2014MNRAS.440.1690H}
{Head}, J.~T.~C.~G., {Lucey}, J.~R., {Hudson}, M.~J., \& {Smith}, R.~J. 2014,
  \mnras, 440, 1690

\bibitem[{{Hilz} {et~al.}(2013){Hilz}, {Naab}, \&
  {Ostriker}}]{2013MNRAS.429.2924H}
{Hilz}, M., {Naab}, T., \& {Ostriker}, J.~P. 2013, \mnras, 429, 2924

\bibitem[{{Hoffman} {et~al.}(2010){Hoffman}, {Cox}, {Dutta}, \&
  {Hernquist}}]{2010ApJ...723..818H}
{Hoffman}, L., {Cox}, T.~J., {Dutta}, S., \& {Hernquist}, L. 2010, \apj, 723,
  818

\bibitem[{{Hopkins} {et~al.}(2009{\natexlab{a}}){Hopkins}, {Cox}, {Dutta},
  {Hernquist}, {Kormendy}, \& {Lauer}}]{2009ApJS..181..135H}
{Hopkins}, P.~F., {Cox}, T.~J., {Dutta}, S.~N., {et~al.} 2009{\natexlab{a}},
  \apjs, 181, 135

\bibitem[{{Hopkins} {et~al.}(2009{\natexlab{b}}){Hopkins}, {Cox}, {Younger}, \&
  {Hernquist}}]{2009ApJ...691.1168H}
{Hopkins}, P.~F., {Cox}, T.~J., {Younger}, J.~D., \& {Hernquist}, L.
  2009{\natexlab{b}}, \apj, 691, 1168

\bibitem[{{Jesseit} {et~al.}(2009){Jesseit}, {Cappellari}, {Naab}, {Emsellem},
  \& {Burkert}}]{2009MNRAS.397.1202J}
{Jesseit}, R., {Cappellari}, M., {Naab}, T., {Emsellem}, E., \& {Burkert}, A.
  2009, \mnras, 397, 1202

\bibitem[{{Kormendy} \& {Bender}(2012)}]{2012ApJS..198....2K}
{Kormendy}, J. \& {Bender}, R. 2012, \apjs, 198, 2

\bibitem[{{Krajnovi{\'c}} {et~al.}(2011){Krajnovi{\'c}}, {Emsellem},
  {Cappellari}, {Alatalo}, {Blitz}, {Bois}, {Bournaud}, {Bureau}, {Davies},
  {Davis}, {de Zeeuw}, {Khochfar}, {Kuntschner}, {Lablanche}, {McDermid},
  {Morganti}, {Naab}, {Oosterloo}, {Sarzi}, {Scott}, {Serra}, {Weijmans}, \&
  {Young}}]{2011MNRAS.414.2923K}
{Krajnovi{\'c}}, D., {Emsellem}, E., {Cappellari}, M., {et~al.} 2011, \mnras,
  414, 2923

\bibitem[{{Laurikainen} {et~al.}(2010){Laurikainen}, {Salo}, {Buta}, {Knapen},
  \& {Comer{\'o}n}}]{2010MNRAS.405.1089L}
{Laurikainen}, E., {Salo}, H., {Buta}, R., {Knapen}, J.~H., \& {Comer{\'o}n},
  S. 2010, \mnras, 405, 1089

\bibitem[{{Maltby} {et~al.}(2015){Maltby}, {Arag{\'o}n-Salamanca}, {Gray},
  {Hoyos}, {Wolf}, {Jogee}, \& {B{\"o}hm}}]{2015MNRAS.447.1506M}
{Maltby}, D.~T., {Arag{\'o}n-Salamanca}, A., {Gray}, M.~E., {et~al.} 2015,
  \mnras, 447, 1506

\bibitem[{{Mapelli} {et~al.}(2015){Mapelli}, {Rampazzo}, \&
  {Marino}}]{2015A&A...575A..16M}
{Mapelli}, M., {Rampazzo}, R., \& {Marino}, A. 2015, \aap, 575, A16

\bibitem[{{Mazzei} {et~al.}(2014{\natexlab{a}}){Mazzei}, {Marino}, \&
  {Rampazzo}}]{2014ApJ...782...53M}
{Mazzei}, P., {Marino}, A., \& {Rampazzo}, R. 2014{\natexlab{a}}, \apj, 782, 53

\bibitem[{{Mazzei} {et~al.}(2014{\natexlab{b}}){Mazzei}, {Marino}, {Rampazzo},
  {Galletta}, \& {Bettoni}}]{2014AdSpR..53..950M}
{Mazzei}, P., {Marino}, A., {Rampazzo}, R., {Galletta}, G., \& {Bettoni}, D.
  2014{\natexlab{b}}, Advances in Space Research, 53, 950

\bibitem[{{Moody} {et~al.}(2014){Moody}, {Romanowsky}, {Cox}, {Novak}, \&
  {Primack}}]{2014MNRAS.444.1475M}
{Moody}, C.~E., {Romanowsky}, A.~J., {Cox}, T.~J., {Novak}, G.~S., \&
  {Primack}, J.~R. 2014, \mnras, 444, 1475

\bibitem[{{Moster} {et~al.}(2011){Moster}, {Macci{\`o}}, {Somerville}, {Naab},
  \& {Cox}}]{2011MNRAS.415.3750M}
{Moster}, B.~P., {Macci{\`o}}, A.~V., {Somerville}, R.~S., {Naab}, T., \&
  {Cox}, T.~J. 2011, \mnras, 415, 3750

\bibitem[{{Naab} \& {Burkert}(2003)}]{2003ApJ...597..893N}
{Naab}, T. \& {Burkert}, A. 2003, \apj, 597, 893

\bibitem[{{Naab} {et~al.}(2014){Naab}, {Oser}, {Emsellem}, {Cappellari},
  {Krajnovi{\'c}}, {McDermid}, {Alatalo}, {Bayet}, {Blitz}, {Bois}, {Bournaud},
  {Bureau}, {Crocker}, {Davies}, {Davis}, {de Zeeuw}, {Duc}, {Hirschmann},
  {Johansson}, {Khochfar}, {Kuntschner}, {Morganti}, {Oosterloo}, {Sarzi},
  {Scott}, {Serra}, {Ven}, {Weijmans}, \& {Young}}]{2014MNRAS.444.3357N}
{Naab}, T., {Oser}, L., {Emsellem}, E., {et~al.} 2014, \mnras, 444, 3357

\bibitem[{{Naab} \& {Trujillo}(2006)}]{2006MNRAS.369..625N}
{Naab}, T. \& {Trujillo}, I. 2006, \mnras, 369, 625 (N06)

\bibitem[{{Querejeta} {et~al.}(2015){Querejeta}, {Eliche-Moral}, {Tapia},
  {Borlaff}, {Rodr{\'{\i}}guez-P{\'e}rez}, {Zamorano}, \&
  {Gallego}}]{2015A&A...573A..78Q}
{Querejeta}, M., {Eliche-Moral}, M.~C., {Tapia}, T., {et~al.} 2015, \aap, 573,
  A78

\bibitem[{{Robertson} {et~al.}(2006){Robertson}, {Hernquist}, {Cox}, {Di
  Matteo}, {Hopkins}, {Martini}, \& {Springel}}]{2006ApJ...641...90R}
{Robertson}, B., {Hernquist}, L., {Cox}, T.~J., {et~al.} 2006, \apj, 641, 90

\bibitem[{{Robertson} {et~al.}(2004){Robertson}, {Yoshida}, {Springel}, \&
  {Hernquist}}]{2004ApJ...606...32R}
{Robertson}, B., {Yoshida}, N., {Springel}, V., \& {Hernquist}, L. 2004, \apj,
  606, 32

\bibitem[{{S{\'a}nchez} {et~al.}(2012){S{\'a}nchez}, {Kennicutt}, {Gil de Paz},
  {van de Ven}, {V{\'{\i}}lchez}, {Wisotzki}, {Walcher}, {Mast}, {Aguerri},
  {Albiol-P{\'e}rez}, {Alonso-Herrero}, {Alves}, {Bakos}, {Bart{\'a}kov{\'a}},
  {Bland-Hawthorn}, {Boselli}, {Bomans}, {Castillo-Morales}, {Cortijo-Ferrero},
  {de Lorenzo-C{\'a}ceres}, {Del Olmo}, {Dettmar}, {D{\'{\i}}az}, {Ellis},
  {Falc{\'o}n-Barroso}, {Flores}, {Gallazzi}, {Garc{\'{\i}}a-Lorenzo},
  {Gonz{\'a}lez Delgado}, {Gruel}, {Haines}, {Hao}, {Husemann},
  {Igl{\'e}sias-P{\'a}ramo}, {Jahnke}, {Johnson}, {Jungwiert}, {Kalinova},
  {Kehrig}, {Kupko}, {L{\'o}pez-S{\'a}nchez}, {Lyubenova}, {Marino},
  {M{\'a}rmol-Queralt{\'o}}, {M{\'a}rquez}, {Masegosa}, {Meidt},
  {Mendez-Abreu}, {Monreal-Ibero}, {Montijo}, {Mour{\~a}o}, {Palacios-Navarro},
  {Papaderos}, {Pasquali}, {Peletier}, {P{\'e}rez}, {P{\'e}rez}, {Quirrenbach},
  {Rela{\~n}o}, {Rosales-Ortega}, {Roth}, {Ruiz-Lara},
  {S{\'a}nchez-Bl{\'a}zquez}, {Sengupta}, {Singh}, {Stanishev}, {Trager},
  {Vazdekis}, {Viironen}, {Wild}, {Zibetti}, \&
  {Ziegler}}]{2012A&A...538A...8S}
{S{\'a}nchez}, S.~F., {Kennicutt}, R.~C., {Gil de Paz}, A., {et~al.} 2012,
  \aap, 538, A8

\bibitem[{{Semelin} \& {Combes}(2002)}]{2002A&A...388..826S}
{Semelin}, B. \& {Combes}, F. 2002, \aap, 388, 826

\bibitem[{{Spitzer} \& {Baade}(1951)}]{1951ApJ...113..413S}
{Spitzer}, Jr., L. \& {Baade}, W. 1951, \apj, 113, 413

\bibitem[{{Springel} {et~al.}(2005){Springel}, {Di Matteo}, \&
  {Hernquist}}]{2005ApJ...620L..79S}
{Springel}, V., {Di Matteo}, T., \& {Hernquist}, L. 2005, \apjl, 620, L79

\bibitem[{{Springel} \& {Hernquist}(2005)}]{2005ApJ...622L...9S}
{Springel}, V. \& {Hernquist}, L. 2005, \apjl, 622, L9

\bibitem[{{Tapia} {et~al.}(2014){Tapia}, {Eliche-Moral}, {Querejeta},
  {Balcells}, {C{\'e}sar Gonz{\'a}lez-Garc{\'{\i}}a}, {Prieto}, {Aguerri},
  {Gallego}, {Zamorano}, {Rodr{\'{\i}}guez-P{\'e}rez}, \&
  {Borlaff}}]{2014A&A...565A..31T}
{Tapia}, T., {Eliche-Moral}, M.~C., {Querejeta}, M., {et~al.} 2014, \aap, 565,
  A31

\bibitem[{{Tsatsi} {et~al.}(2015){Tsatsi}, {Macci{\`o}}, {van de Ven}, \&
  {Moster}}]{2015ApJ...802L...3T}
{Tsatsi}, A., {Macci{\`o}}, A.~V., {van de Ven}, G., \& {Moster}, B.~P. 2015,
  \apjl, 802, L3

\bibitem[{{Vaghmare} {et~al.}(2015){Vaghmare}, {Barway}, {Mathur}, \&
  {Kembhavi}}]{2015arXiv150307635V}
{Vaghmare}, K., {Barway}, S., {Mathur}, S., \& {Kembhavi}, A.~K. 2015, ArXiv
  e-prints

\bibitem[{{van den Bergh}(1976)}]{1976ApJ...206..883V}
{van den Bergh}, S. 1976, \apj, 206, 883

\bibitem[{{Vijayaraghavan} \& {Ricker}(2013)}]{2013MNRAS.435.2713V}
{Vijayaraghavan}, R. \& {Ricker}, P.~M. 2013, \mnras, 435, 2713

\bibitem[{{Weil} \& {Hernquist}(1996)}]{1996ApJ...460..101W}
{Weil}, M.~L. \& {Hernquist}, L. 1996, \apj, 460, 101

\bibitem[{{Wilman} {et~al.}(2009){Wilman}, {Oemler}, {Mulchaey}, {McGee},
  {Balogh}, \& {Bower}}]{2009ApJ...692..298W}
{Wilman}, D.~J., {Oemler}, Jr., A., {Mulchaey}, J.~S., {et~al.} 2009, \apj,
  692, 298

\bibitem[{{Zinchenko} {et~al.}(2015){Zinchenko}, {Berczik}, {Grebel},
  {Pilyugin}, \& {Just}}]{2015arXiv150407483Z}
{Zinchenko}, I.~A., {Berczik}, P., {Grebel}, E.~K., {Pilyugin}, L.~S., \&
  {Just}, A. 2015, ArXiv e-prints

\end{thebibliography}

\clearpage
\onecolumn

{
\footnotesize
\begin{center}

\begin{minipage}[t]{16cm}
\vspace{0.5cm}
\end{minipage}
  \begin{longtable}{c l c c c  c c c   c c c}
\caption{Properties of the simulated progenitors and remnants.}
\label{tab:parameters}
\\ \hline \\\vspace{-0.6cm}\\
No. & \multicolumn{1}{c}{Model} & Spin  & Inclination & Pericentre & Velocity & Mass ratio & Gas fraction  &  $\varepsilon_\mathrm{e}$  & $R_{90}/R_{50}$  & $\lambda_\mathrm{Re}$  \\
 &   &   & [deg] & [kpc] & [km/s] &  & \% &   &   &  \\ 
(1) & \multicolumn{1}{c}{(2)} & (3)  & (4) & (5)  & (6)  & (7) & (8) & (9) & (10) & (11)  \vspace{0.1cm}\\\hline
\\\vspace{-0.5cm}\\
\endhead
 -- & Original gSa & -- & -- & -- & -- & -- & -- &    0.87 &    2.70 &    0.79\\
 -- & Original gSb & -- & -- & -- & -- & -- & -- &    0.88 &    3.21 &    0.78\\
 -- & Original gSd & -- & -- & -- & -- & -- & -- &    0.91 &    2.21 &    0.91\\
 \hline\\\vspace{-0.5cm}\\
  1 & gE0gSao1 & P &   0 &   8 &  200 & 3:2 &   3 &     0.37 &     3.71 &     0.30 \\ 
  2 & gE0gSao5 & P &   0 &  16 &  200 & 3:2 &   3  &     0.41 &     3.09 &     0.37 \\ 
  3 & gE0gSao16 & R &   0 &   8 &  200 & 3:2 &   3  &     0.52 &     3.53 &     0.16 \\ 
  4 & gE0gSao44 & R &   0 &  16 &  200 & 3:2 &   3  &     0.47 &     3.63 &     0.26 \\ 
  5 & gE0gSbo5 & P &   0 &  16 &  200 & 3:1 &   4  &     0.35 &     3.09 &     0.24 \\ 
  6 & gE0gSbo44 & R &   0 &  16 &  200 & 3:1 &   4  &     0.33 &     3.38 &     0.27 \\ 
  7 & gE0gSdo5 & P &   0 &  16 &  200 & 3:1 &   7  &     0.31 &     3.42 &     0.25 \\ 
  8 & gE0gSdo16 & R &   0 &   8 &  200 & 3:1 &   7  &     0.28 &     2.90 &     0.20 \\ 
  9 & gE0gSdo44 & R &   0 &  16 &  200 & 3:1 &   7  &     0.28 &     3.64 &     0.25 \\ 
 10 & gSagSao1 & P &   0 &   8 &  200 & 1:1 &   8  &     0.75 &     3.83 &     0.43 \\ 
 11 & gSagSao5 & P &   0 &  16 &  200 & 1:1 &   8  &     0.67 &     4.18 &     0.52 \\ 
 12 & gSagSao9 & P &   0 &  24 &  200 & 1:1 &   8  &     0.61 &     3.33 &     0.59 \\ 
 13 & gSagSbo1 & P &   0 &   8 &  200 & 2:1 &  10  &     0.73 &     3.82 &     0.43 \\ 
 14 & gSagSbo2 & P &   0 &   8 &  300 & 2:1 &  10  &     0.68 &     3.84 &     0.35 \\ 
 15 & gSagSbo5 & P &   0 &  16 &  200 & 2:1 &  10  &     0.65 &     3.85 &     0.52 \\ 
 16 & gSagSbo9 & P &   0 &  24 &  200 & 2:1 &  10  &     0.61 &     3.39 &     0.54 \\ 
 17 & gSagSbo21 & P &  75 &   8 &  300 & 2:1 &  10  &     0.56 &     2.97 &     0.35 \\ 
 18 & gSagSbo22 & P &  90 &   8 &  300 & 2:1 &  10  &     0.48 &     3.52 &     0.34 \\ 
 19 & gSagSbo24 & R &  45 &   8 &  300 & 2:1 &  10  &     0.47 &     3.05 &     0.09 \\ 
 20 & gSagSbo42 & P &  75 &  16 &  200 & 2:1 &  10  &     0.55 &     3.56 &     0.53 \\ 
 21 & gSagSbo43 & P &  90 &  16 &  200 & 2:1 &  10  &     0.51 &     3.70 &     0.51 \\ 
 22 & gSagSbo71 & P &  90 &  24 &  200 & 2:1 &  10  &     0.52 &     3.45 &     0.61 \\ 
 23 & gSagSdo2 & P &   0 &   8 &  300 & 2:1 &  15  &     0.64 &     4.80 &     0.45 \\ 
 24 & gSagSdo5 & P &   0 &  16 &  200 & 2:1 &  15  &     0.70 &     3.86 &     0.54 \\ 
 25 & gSagSdo9 & P &   0 &  24 &  200 & 2:1 &  15  &     0.66 &     4.44 &     0.52 \\ 
 26 & gSagSdo14 & P &  75 &   8 &  200 & 2:1 &  15  &     0.59 &     3.67 &     0.48 \\ 
 27 & gSagSdo18 & R &  75 &   8 &  200 & 2:1 &  15  &     0.09 &     3.06 &     0.07 \\ 
 28 & gSagSdo41 & P &  45 &  16 &  200 & 2:1 &  15  &     0.56 &     3.46 &     0.56 \\ 
 29 & gSagSdo42 & P &  75 &  16 &  200 & 2:1 &  15  &     0.54 &     3.06 &     0.54 \\ 
 30 & gSagSdo43 & P &  90 &  16 &  200 & 2:1 &  15  &     0.53 &     3.11 &     0.52 \\ 
 31 & gSagSdo70 & P &  75 &  24 &  200 & 2:1 &  15  &     0.53 &     3.93 &     0.61 \\ 
 32 & gSagSdo71 & P &  90 &  24 &  200 & 2:1 &  15  &     0.53 &     4.38 &     0.62 \\ 
 33 & gSagSdo73 & R &  45 &  24 &  200 & 2:1 &  15  &     0.08 &     4.61 &     0.11 \\ 
 34 & gSbgSbo9 & P &   0 &  24 &  200 & 1:1 &  16  &     0.69 &     3.31 &     0.48 \\ 
 35 & gSbgSbo16 & R &   0 &   8 &  200 & 1:1 &  16  &     0.57 &     3.48 &     0.10 \\ 
 36 & gSbgSbo17 & R &  45 &   8 &  200 & 1:1 &  16  &     0.27 &     2.81 &     0.27 \\ 
 37 & gSbgSbo19 & R &  90 &   8 &  200 & 1:1 &  16  &     0.25 &     3.19 &     0.11 \\ 
 38 & gSbgSbo22 & P &  90 &   8 &  300 & 1:1 &  16  &     0.49 &     3.64 &     0.15 \\ 
 39 & gSbgSbo41 & P &  45 &  16 &  200 & 1:1 &  16  &     0.34 &     2.96 &     0.42 \\ 
 40 & gSbgSbo42 & P &  75 &  16 &  200 & 1:1 &  16  &     0.41 &     3.80 &     0.46 \\ 
 41 & gSbgSbo69 & P &  45 &  24 &  200 & 1:1 &  16  &     0.35 &     2.60 &     0.56 \\ 
 42 & gSbgSbo70 & P &  75 &  24 &  200 & 1:1 &  16  &     0.46 &     3.63 &     0.59 \\ 
 43 & gSbgSbo72 & R &   0 &  24 &  200 & 1:1 &  16  &     0.55 &     4.24 &     0.17 \\ 
 44 & gSbgSdo5 & P &   0 &  16 &  200 & 1:1 &  23  &     0.66 &     4.88 &     0.50 \\ 
 45 & gSbgSdo9 & P &   0 &  24 &  200 & 1:1 &  23  &     0.68 &     3.38 &     0.52 \\ 
 46 & gSbgSdo14 & P &  75 &   8 &  200 & 1:1 &  23  &     0.35 &     3.03 &     0.35 \\ 
 47 & gSbgSdo17 & R &  45 &   8 &  200 & 1:1 &  23  &     0.47 &     2.96 &     0.36 \\ 
 48 & gSbgSdo18 & R &  75 &   8 &  200 & 1:1 &  23  &     0.26 &     4.14 &     0.25 \\ 
 49 & gSbgSdo19 & R &  90 &   8 &  200 & 1:1 &  23  &     0.30 &     3.28 &     0.29 \\ 
 50 & gSbgSdo23 & R &   0 &   8 &  300 & 1:1 &  23  &     0.42 &     3.09 &     0.18 \\ 
 51 & gSbgSdo41 & P &  45 &  16 &  200 & 1:1 &  23  &     0.45 &     2.91 &     0.43 \\ 
 52 & gSbgSdo69 & P &  45 &  24 &  200 & 1:1 &  23  &     0.42 &     2.77 &     0.47 \\ 
 53 & gSbgSdo70 & P &  75 &  24 &  200 & 1:1 &  23  &     0.49 &     3.10 &     0.55 \\ 
 54 & gSbgSdo71 & P &  90 &  24 &  200 & 1:1 &  23  &     0.39 &     3.18 &     0.56 \\ 
 55 & gSdgSdo1 & P &   0 &   8 &  200 & 1:1 &  30  &     0.54 &     5.55 &     0.62 \\ 
 56 & gSdgSdo2 & P &   0 &   8 &  300 & 1:1 &  30  &     0.67 &     0.54 &     0.32 \\ 
 57 & gSdgSdo5 & P &   0 &  16 &  200 & 1:1 &  30  &     0.67 &     5.40 &     0.51 \\ 
 58 & gSdgSdo9 & P &   0 &  24 &  200 & 1:1 &  30  &     0.69 &     3.50 &     0.50 \\ 
 59 & gSdgSdo16 & R &   0 &   8 &  200 & 1:1 &  30  &     0.46 &     3.16 &     0.08 \\ 
 60 & gSdgSdo17 & R &  45 &   8 &  200 & 1:1 &  30  &     0.36 &     3.03 &     0.45 \\ 
 61 & gSdgSdo21 & P &  75 &   8 &  300 & 1:1 &  30  &     0.25 &     3.01 &     0.37 \\ 
 62 & gSdgSdo42 & P &  75 &  16 &  200 & 1:1 &  30  &     0.48 &     4.85 &     0.46 \\ 
 63 & gSdgSdo45 & R &  45 &  16 &  200 & 1:1 &  30  &     0.06 &     3.13 &     0.29 \\ 
 64 & gSdgSdo51 & R &   0 &  16 &  300 & 1:1 &  30  &     0.06 &     5.06 &     0.18 \\ 
 65 & gSdgSdo69 & P &  45 &  24 &  200 & 1:1 &  30  &     0.47 &     3.06 &     0.65 \\ 
 66 & gSdgSdo71 & P &  90 &  24 &  200 & 1:1 &  30  &     0.28 &     2.98 &     0.50 \\ 
 67 & gSdgSdo74 & R &  75 &  24 &  200 & 1:1 &  30  &     0.37 &     3.50 &     0.33 \\ 
 \hline\\
 \vspace{-0.6cm}
 \end{longtable} 
\end{center}
}
\centering
\begin{minipage}[t]{16cm}
{\footnotesize
\emph{Columns}: (1) ID number. (2) GalMer model code: g[\emph{type1}]g[\emph{type2}]o[\emph{\#orbit}]. (3) Spin-orbit coupling of the encounter (P, prograde; R, retrograde). (4) Initial inclination between the planes of the discs involved in the merger. (5) Pericentre distance of the initial orbit, in kpc. (6) Absolute value of the initial relative velocity, in km/s. (7) Stellar mass ratio. (8) Initial gas fraction involved in the encounter. (9) Ellipticity of the remnant ($\varepsilon_\mathrm{e}$) measured in edge-on view. (10) Concentration of the remnant, $R_{90}/R_{50}$. (11) Angular momentum of the remnant ($\lambda_\mathrm{Re}$) measured in edge-on view.
}
\end{minipage}

\clearpage
\twocolumn
\end{document}